\documentclass[a4paper,11pt]{article}
\pdfoutput=1 

\usepackage{jcappub} 

\usepackage[T1]{fontenc} 
\usepackage{caption,subcaption}
\usepackage{epsfig}
\usepackage{epstopdf}
\title{Massive white dwarfs in Rastall-Rainbow gravity}

\author[a]{Jie Li,}
\author[b]{Bo Yang,}
\author[b,c,1]{Wenbin Lin,\note{Corresponding author.}}

\affiliation[a]{School of Nuclear Science and Technology, \\University of South China, Hengyang 421001, China}
\affiliation[b]{School of Mathematics and Physics, \\University of South China, Hengyang, 421001, China}
\affiliation[c]{School of Physical Science and Technology,\\Southwest Jiaotong University, Chengdu, 610031, China}

\emailAdd{lij@stu.usc.edu.cn}
\emailAdd{yb@usc.edu.cn}
\emailAdd{lwb@usc.edu.cn}

\abstract{We investigate the hydrostatic equilibrium of white dwarfs within the framework of Rastall-Rainbow gravity, aiming to explore the effects of this modified gravitational theory on their properties. By employing the Chandrasekhar equation of state in conjunction with the modified Tolman-Oppenheimer-Volkoff equation, we derive the mass-radius relations for white dwarfs. Our results show that the maximum mass of white dwarfs deviates significantly from the predictions of general relativity, potentially exceeding the Chandrasekhar limit. Furthermore, we discuss other properties of white dwarfs, such as the gravitational redshift, compactness and dynamical stability, shedding light on their behavior within the context of this modified gravitational framework.}

\begin{document}
\maketitle
\flushbottom

\section{Introduction}
Einstein's general relativity (GR), proposed a century ago, is a successful theory of gravity which has helped us understand many aspects of the universe. In particular, some experimental tests from weak gravity, as well as recent  observations like the shadows of black holes and gravitational waves by binary stars mergers, have strongly supported GR. Nonetheless, there are some phenomena at the  larger (cosmological) scale that are not well described in the context of GR. For example, GR fails to explain the accelerated expansion of the universe without further refinement. To overcome this  limitation, two possibilities were  proposed.  One is that there is a large amount of dark energy with negative pressure in the universe. The other is that GR's predictions may be biased on the cosmological scale, so alternative theories of gravity were proposed, such as massive gravity~\cite{Rham2011}, Brans-Dicke gravity~\cite{Brans1961} and $f(R)$ gravity~\cite{Felice2010}.

Additionally, Rastall~\cite{Rastall1972} proposed a possible modified theory of gravity. Rastall argued that current experiments merely verified the vanishing of the covariant derivative of the energy-momentum tensor in flat spacetimes, without necessarily holding true in curved spacetime. Thus, Rastall replaced the energy-momentum tensor conservation law in GR by assuming that the covariant divergence of the energy-momentum tensor varies proportionally to the gradient of the Ricci scalar. In the past few years, Rastall gravity has attracted attention from researchers  due to its potential to explain some phenomena in the universe~\cite{Batista2012,Fabris2012,Moradpour2017,Darabi2018}.   Recently, several works have been done to test  Rastall gravity by studying compact stars with strong gravitational fields. Some solutions of black hole in the Rastall gravity are obtained in Refs.~\cite{Bronnikov2016,Heydarzade2017,Ma2017,Heydarzade20172,Graca2018,Xu2018}, as well as various physical properties of these black holes have been extensively studied~\cite{Bamba2018,Lobo2018}. Oliveira et al.,~\cite{Oliveira2015} first obtained the modified equilibrium equation in this gravity, and discussed the effects of Rastall parameter on the structure of neutron stars. Ghosh et al.,~\cite{Ghosh2021} investigated gravastars in Rastall gravity. In the literatures~\cite{Bhar2020,Abbas2019}, the authors studied anisotropic compact stars in Rastall gravitational framework using Karori-Barua  metric or Tolman-Kuchowicz metric. Quark stars and quintessence compact stars were discussed in Refs.~\cite{Abbas2018,Shahzad2019}.

Rainbow gravity was proposed by Magueijo and Smolin~\cite{Magueijo2004}, who incorporate double special relativity into the framework of GR. Recently, Mota et al.~\cite{Mota2019}combined Rastall gravity with Rainbow theory and employed the modern equation of state (EoS) of nuclear matter to derive models of neutron stars. They concluded that slight changes in both parameters of this theory of gravity  have a substantial impact on the structure of the neutron star. Subsequently, anisotropic neutron stars in this gravitational background was studied~\cite{Mota2022}. Charged anisotropic strange stars and charged gravastars in Rastall-Rainbow gravity were investigated in Refs.~\cite{Debnath2021,Das2022}.

White dwarfs are extraordinarily dense celestial bodies that formed from  the collapse of stellar stars following the complete depletion of their nuclear fuel within the central region. The study of white dwarfs has been crucial in advancing our understanding of stellar evolution and the physics of compact objects. In the past few years, a lot of researches have been conducted on the relationship between the mass and radius of white dwarfs, revealing that  the existence of an upper mass limit known as the Chandrasekhar mass limit~\cite{Chandrasekhar1931,Chandrasekhar1935}. However, some recently observed peculiar over-luminous type-Ia supernovae, such as SN 2007if, SN 2006gz, SN 2003fg and SN 2009DC~\cite{Howell2006,Scalzo2010}, predicted the existence of white dwarfs with masses ranging from 2.1 $M_{\odot}$  to 2.8 $M_{\odot}$ ($M_{\odot}$ is the mass of the Sun)~\cite{Hicke2007,Yamanaka2009,Silverman2011,Taubeberger2011}. In order to explain the formation mechanism of super-Chandrasekhar white dwarfs, extensive research investigated white dwarfs in various contexts, such as super strong uniform magnetized white dwarfs~\cite{Das2012,Das2013,Das2014,Franzon2015,Deb2022}, white dwarfs in modified theories of gravity~\cite{Das2015,Jing2016,Banerjee2017,Carvalho17,Kalita2018,Panah2019,Liu2019,Rocha2020,Wojnar2021,Kalita2021,Kalita2022}, electrical charged white dwarfs~\cite{Liu2014,Carvalho2018} and rotating white dwarfs~\cite{Boshkayev2011,Boshkayev2012}. In this work, we explore the behavior of white dwarfs within the framework of Rastall-Rainbow gravity, specifically investigating the potential existence of super-Chandrasekhar white dwarfs within this modified gravitational theory.

This article is organized as follows. In Sec. II, we briefly review Rastall gravity and Rainbow gravity, and proceed to derive the modified Tolman-Oppenheimer-Volkoff (TOV) equation within the framework of Rastall-Rainbow gravity. In Sec. III, we recall the Chandrasekhar EoS. In Sec. IV, we solve modified TOV equation numerically and present the mass-radius relations for white dwarfs in Rastall-Rainbow gravity. In Sec. V, some properties of white stars are investigated, including the gravitational redshift, compactness and dynamic stability. Finally, summary is given in Sec. VI.

\section{Basic equations of Rastall-Rainbow gravity}
\subsection{Rastall-Rainbow gravity}
Double special relativity was proposed due to the existence of the minimum observable length, which requires the speed of light and the Planck energy remain constant in the low energy limit. Consequently, the energy-momentum dispersion relation for a particle of mass $m$ can be written in the following general form
\begin{eqnarray}
E^{2}\Delta^{2}(\varepsilon)-p^{2}\Theta^{2}(\varepsilon)=m^{2}~,\label{dispersion-relation}
\end{eqnarray}
where $\varepsilon = E/E_{p}$ is the ratio of the particle's energy $E$ to the Planck constant $E_{p}$. $p$ is  the particle's momentum. $\Delta(\varepsilon)$ and $\Theta(\varepsilon)$ are known as the rainbow functions, and they satisfy the conditions
\begin{eqnarray}
\lim_{\varepsilon\rightarrow0}~\Delta(\varepsilon)=1,~~~~ \lim_{\varepsilon\rightarrow0}~\Theta(\varepsilon)=1~;\label{rainbow}
\end{eqnarray}
Magueijo and Smolin integrated the double special relativity into the framework of general relativity to propose Rainbow gravity. In this proposal, the modified metric can be written as
\begin{eqnarray}
g(\varepsilon)=\eta^{ab}e_{a}(\varepsilon) \tiny \otimes e_{b}(\varepsilon)~,\label{metric}
\end{eqnarray}
where $e_{a}(\varepsilon)$ and $e_{b}(\varepsilon)$  are the energy-dependent orthonormal frame fields, and their expressions are as follows
\begin{eqnarray}
e_{0}(\varepsilon)=\frac{1}{\Delta(\varepsilon)}~{e_{0}},~~~ e_{i}(x)=\frac{1}{\Theta(\varepsilon)}~{e_{i}}~;\label{orthonormal}
\end{eqnarray}
According to the above description, the spherical symmetry metric in Rainbow gravity can be written as
\begin{eqnarray}
ds^{2}=-\frac{e^{2\nu}}{\Delta(\varepsilon)^{2}}~dt^{2}+\frac{e^{2\lambda}}{\Theta(\varepsilon)^{2}}~dr^{2}+\frac{r^{2}}{\Theta(\varepsilon)^{2}}~(d\theta^{2}+\sin\theta^{2}d\phi^{2})~,\label{ds}
\end{eqnarray}
where $e^{2\nu}$ and $e^{2\lambda}$ are radial functions. Obviously, the spherically symmetric metric is related to the rainbow function, so it depends on the energy of the probe particle.

According to Rastall's proposal, the conservation law for the energy-momentum tensor may not hold true in curved spacetime, and it should be rewritten as
\begin{eqnarray}
\nabla_{\mu}T^{\mu\nu}= \frac{1-\kappa}{16\pi G} \nabla_{\nu} R~,\label{D-Tuv}
\end{eqnarray}
where  $\kappa$ is a constant called Rastall parameter, which measures the deviation from GR and describes the affinity of the matter field to couple with geometry. The field equation in Rastall's gravity can be written as
\begin{eqnarray}
R_{\mu\nu}-\frac{\kappa}{2} g_{\mu\nu} R=8\pi G T_{\mu\nu}~;\label{field-equations}
\end{eqnarray}
\subsection{Hydrostatic equilibrium equation in Rastall-Rainbow gravity}
Mota et al [8] first combined the above two modified gravitational theories and derived the Hydrostatic equilibrium equations. Next, we introduce the modified TOV equation in this gravitational theory. The field equations in Rastall-Rainbow is
\begin{eqnarray}
R_{\mu\nu}(\varepsilon)-\frac{\kappa}{2}g_{\mu\nu}R(\varepsilon)=8\pi G(\varepsilon)T_{\mu\nu}(\varepsilon)~;\label{Ruv}
\end{eqnarray}
In this investigation, we consider the isotropic perfect fluid whose energy-momentum tensor is
\begin{eqnarray}
T_{\mu\nu}=(P+c^{2}\rho)U_{\mu}U_{\nu}-Pg_{\mu\nu}~,\label{Tuv}
\end{eqnarray}
where $P$ and $\rho$ denote the pressure and the energy density of the fluid, respectively, while  $U_{\mu}$ corresponds to the four-velocity.

By employing Eqs.~\eqref{ds} and~\eqref{Tuv},  we can express the components of the energy-momentum tensor in the four-dimensional spacetime as
\begin{eqnarray}
T^{0}_{0}=\rho c^{2},~~~T^{1}_{1}=T^{2}_{2}=T^{3}_{3}=-P~;\label{energy-momentum}
\end{eqnarray}

Using the energy-momentum tensor Eq.~\eqref{energy-momentum} and the spherical symmetry metric Eq.~\eqref{ds}, the non-vanishing components of the field equations are given as
\begin{eqnarray}
&&\frac{e^{-2\lambda}}{r^{2}}(2r\lambda^{'}-1)+\frac{1}{r^{2}}=\frac{8\pi \rho_{eff}(r)}{c^{2}}~,\label{field-equations-1}\\
&&\frac{e^{-2\lambda}}{r^{2}}(2r\upsilon^{'}+1)-\frac{1}{r^{2}}=\frac{8\pi P_{eff}(r)}{c^{4}}~,\label{field-equations-2}\\
&&\frac{e^{-2\lambda}}{r}\big(\upsilon^{'}+r\upsilon^{'2}-\lambda^{'}-r \lambda^{'} \upsilon^{'}+r\upsilon^{''}\big)=\frac{8\pi P_{eff}(r)}{c^{4}}~,\label{field-equations-3}
\end{eqnarray}
where prime  denotes the differentiation with $r$.  $\rho_{eff}(r)$ and $p_{eff}(r)$  are the effective density and pressure, the expressions are
\begin{eqnarray}
&&\rho_{eff}(r)=\frac{1}{\Delta(x)^2}\Big(\alpha c^{2} \rho+3 \beta P \Big)~,\label{ro-eff}\\
&&P_{eff}(r)=\frac{1}{\Delta(x)^2}\Big(\beta c^{2} \rho+(1-3\beta) P\Big)~,\label{p-eff}
\end{eqnarray}
where
\begin{eqnarray}
\alpha=\frac{2+3\chi}{2~(2+2\chi)},~~~ \beta=\frac{\chi}{2~(2+2\chi)}~;\label{alpha-beta}
\end{eqnarray}
Here we assumed that $\chi = \kappa-1$. The differential equations Eqs.~\eqref{field-equations-1}-\eqref{field-equations-3} are similar to those obtained in GR.  Thus, the expression of the function $e^{-2\lambda}$ can be obtained by directly integrating Eq.~\eqref{field-equations-1}, as
\begin{eqnarray}
e^{-2\lambda}=1-\frac{2Gm(r)}{c^{2}r}~,\label{e-lambda}
\end{eqnarray}
where $m(r)$ is the gravitational mass within the sphere of radius $r$. An additional equation related to the metric function $\nu(r)$ is obtained from Eq.~\eqref{D-Tuv} and reads
\begin{eqnarray}
\frac{d\nu}{dr}=-\Big[\beta\frac{d\rho}{dP}+1-3\beta \Big]\Big[(\alpha+\beta)c^{2}\rho+P \Big]^{-1}\frac{dP}{dr}~;\label{dvdr}
\end{eqnarray}
The expression for $d\nu/dr$ is derived from Eqs.~\eqref{field-equations-1} and~\eqref{e-lambda}, and it is brought into Eq.~\eqref{dvdr}, one can obtain the modified TOV equations in Rastall-Rainbow gravity as
\begin{eqnarray}
&&\frac{dP}{dr}=-\frac{G m}{c^{2}r^{2}}\Big[\Big(\beta\frac{d\rho}{dP}+1-3\beta \Big) \Big(1-\frac{2m}{c^{2}r} \Big)\Big]^{-1}\nonumber\\
&&~~~~~~~~~~~~\Big[\big(\alpha+\beta\big)c^{2}\rho+P \Big]\Big[1+\frac{4\pi r^{3}}{m \Delta^{2}}\Big(\beta c^{2} \rho+P-3\beta P\Big)\Big]~,\label{TOV}
\end{eqnarray}
and
\begin{eqnarray}
\frac{d m(r)}{dr}=\frac{4\pi r^{2}}{\Delta^{2}}\Big(\alpha c^{2} \rho+3\beta P\Big)~;\label{dMdr}
\end{eqnarray}
Naturally, by setting the parameter $\chi = 0$,  that is, $\alpha = 1$ and $\beta = 0$, the modified TOV equation in Rainbow is recovered. When $\chi = 0$ and $\Delta=0$, the standard TOV equation in GR is recovered.
\section{Equation of state}
Solving the modified TOV equation necessitates the adoption of a specific EoS that establishes the interrelation between the pressure and density within the star. White dwarfs, as compact celestial bodies primarily composed of degenerate electron gas, have been extensively studied using the Chandrasekhar EoS~\cite{Chandrasekhar1935}, which can be expressed in the following form
\begin{eqnarray}
&&P(k_{F})=\frac{1}{3\pi^{2}\hbar^{3}}\int^{k_{F}}_{0} \frac{k^{4}}{\sqrt{k^{2}+m_{e}^{2}}}dk \nonumber\\
&&~~~~~~~~~~~~~ =\frac{\pi m_{e}^{4}}{3 h^{3}}\Big[x_{F}(2x_{F}^{2}-3)\sqrt{x_{F}^{2}+1}+3\sinh^{-1}x_{F}\Big]~,\label{Eos-P}\\
&&~~~\rho=\frac{8\pi\mu_{e}m_{H}m_{e}^{3}}{3h^{3}}x_{F}^{3}~,\label{Eos-ro}
\end{eqnarray}
where $x_{F} \equiv p_{F}/m_{e}c$, $p_{F}$ being the Fermi momentum,  $k$ is the momentum of electrons, $m_{e}$ represents the electron mass, $m_{H}$ is the nucleon mass, $h$ denotes the Planck constant.  $\mu_{e}$ is understood as the mean molecular weight per electron. For carbon-oxygen white dwarfs, $\mu_{e} = 2$.
\section{Numerical results}
Some recently observed  peculiar over-luminous SNIa predict the mass of white dwarfs to be between 2.1$M_{\odot}$ and 2.8 $M_{\odot}$. Therefore, the upper limit of the white dwarf's mass is still an open question. In the following, we delve into the properties of white dwarfs within the context of Rastall-Rainbow gravity. The objective is to determine whether this gravitational model can account for the existence of super-Chandrasekhar white dwarfs. In this section, the results for white dwarfs are presented using the Chandrasekhar EoS given in the previous subsection. The mass and radius of white stars are obtained by numerically integrating the  modified TOV equations Eqs.~\eqref{TOV} and~\eqref{dMdr}. Here, we set  the boundary conditions at the center of the star $(r=0)$ as
\begin{eqnarray}
\rho(0)=\rho_{c},~~and~~M(0)=0~,\label{initial}
\end{eqnarray}
where $\rho_{c}$ is the central energy density. The integral is interrupted on the surface of the white dwarf, where the pressure disappears ($P(R)=0$), the corresponding $R$ and $M(R)$ are the radius and total mass of the white dwarf.
\begin{figure}[t]
\centering
\begin{subfigure}[t]{0.49\textwidth}
\centering
\includegraphics[width=8.3cm,height = 6cm]{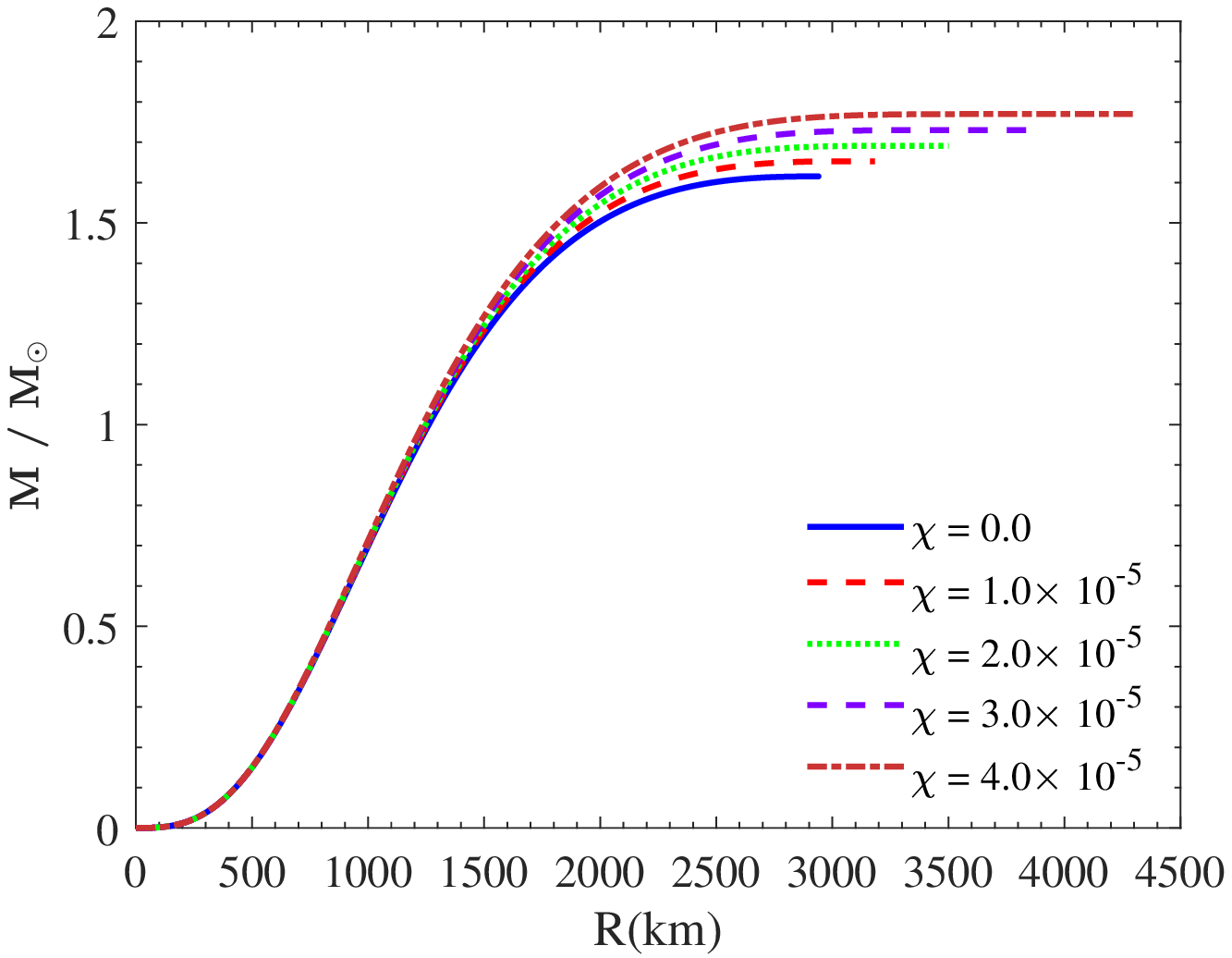}
\end{subfigure}
\begin{subfigure}[t]{0.50\textwidth}
\centering
\includegraphics[width=8.3cm,height = 6cm]{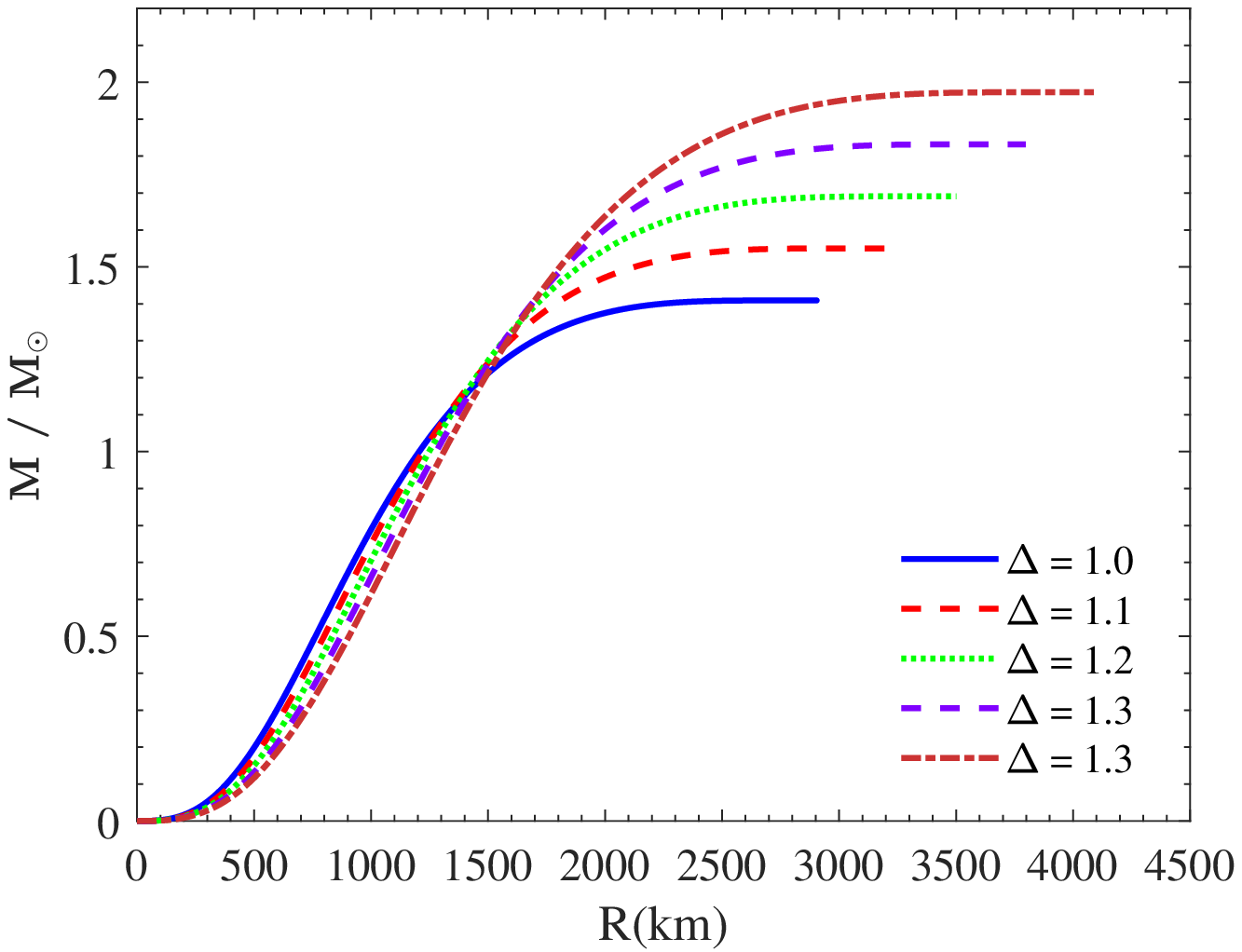}
\end{subfigure}
\caption{Mass function $m(r)$ from the center of white dwarfs to the surface for different parameters $\chi$ (left panel) and $\Delta$ (right panel), where we set $\rho_{c} = 1\times10^{12}~$kg/m$^{3}$.}
\label{M-R}
\end{figure}
\begin{figure}[t]
\centering
\begin{subfigure}[t]{0.49\textwidth}
\centering
\includegraphics[width=8.3cm,height = 6cm]{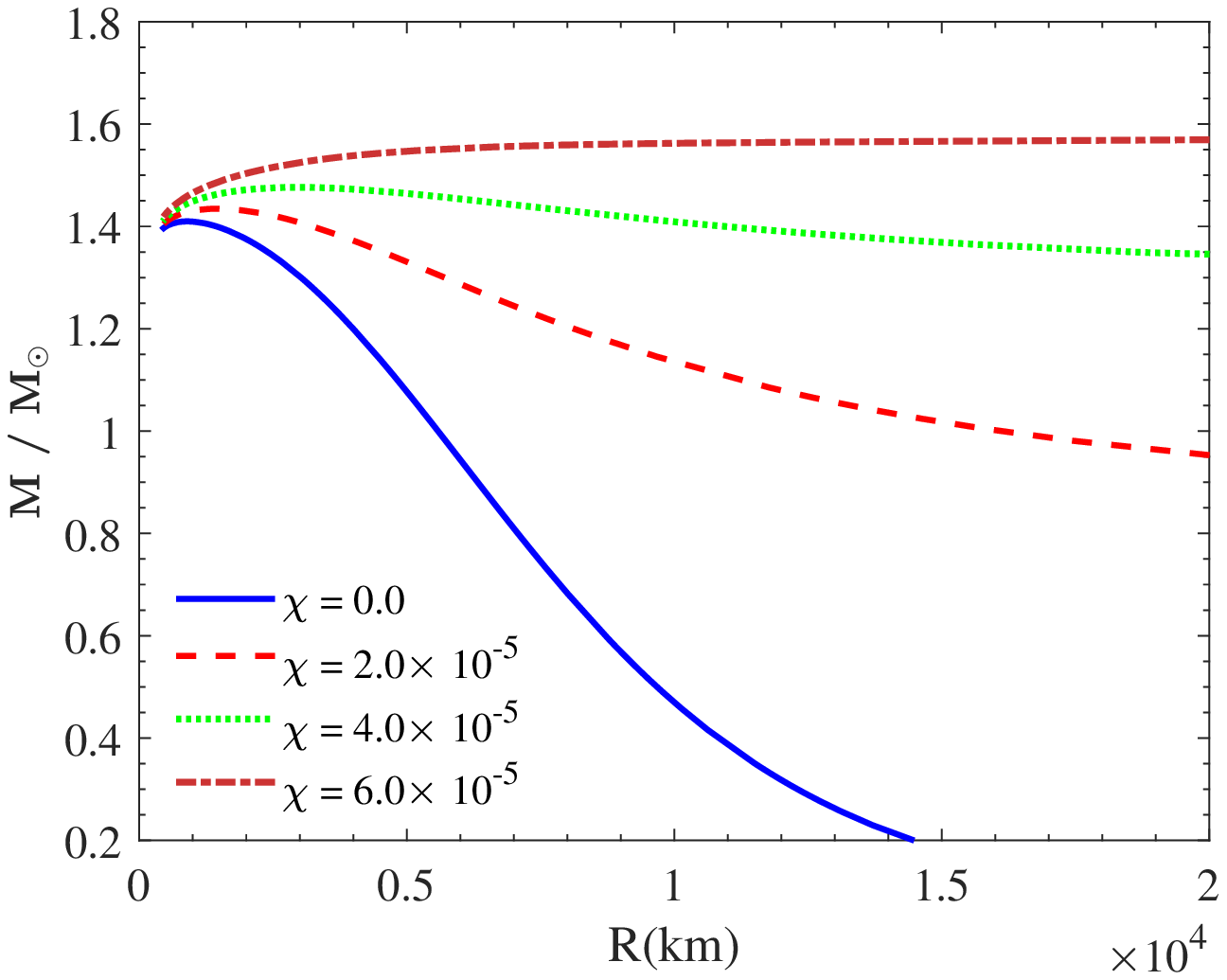}
\end{subfigure}
\begin{subfigure}[t]{0.50\textwidth}
\centering
\includegraphics[width=8.3cm,height = 6cm]{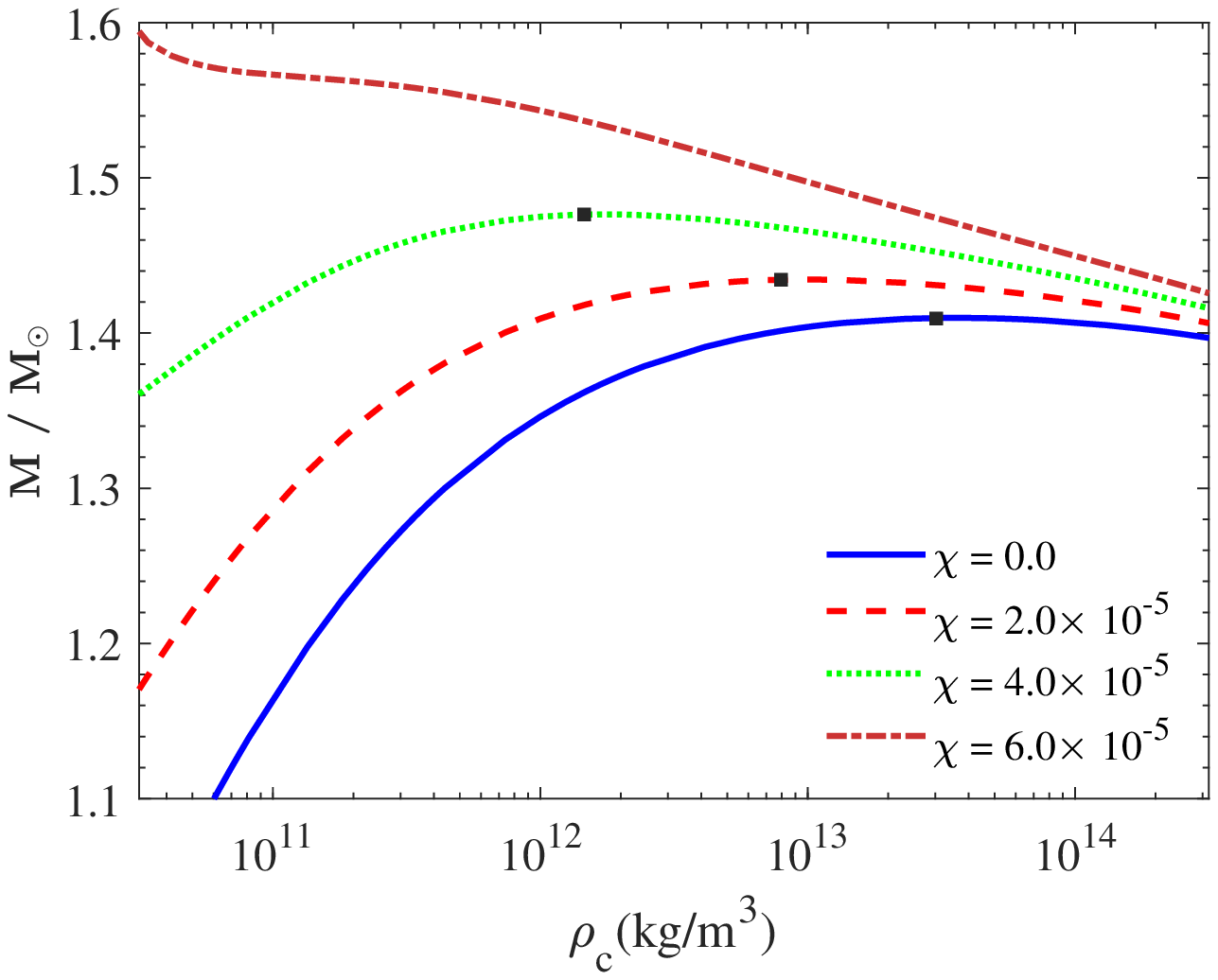}
\end{subfigure}
\caption{The mass-radius (left) and mass-central density $\rho_{c}$  relations (right) for the white stars in Rastall gravity for different values of $\chi$.}
\label{Rastall-M-R-ro}
\end{figure}
\begin{figure}[t]
\centering
\begin{subfigure}[t]{0.49\textwidth}
\centering
\includegraphics[width=8.3cm,height = 6cm]{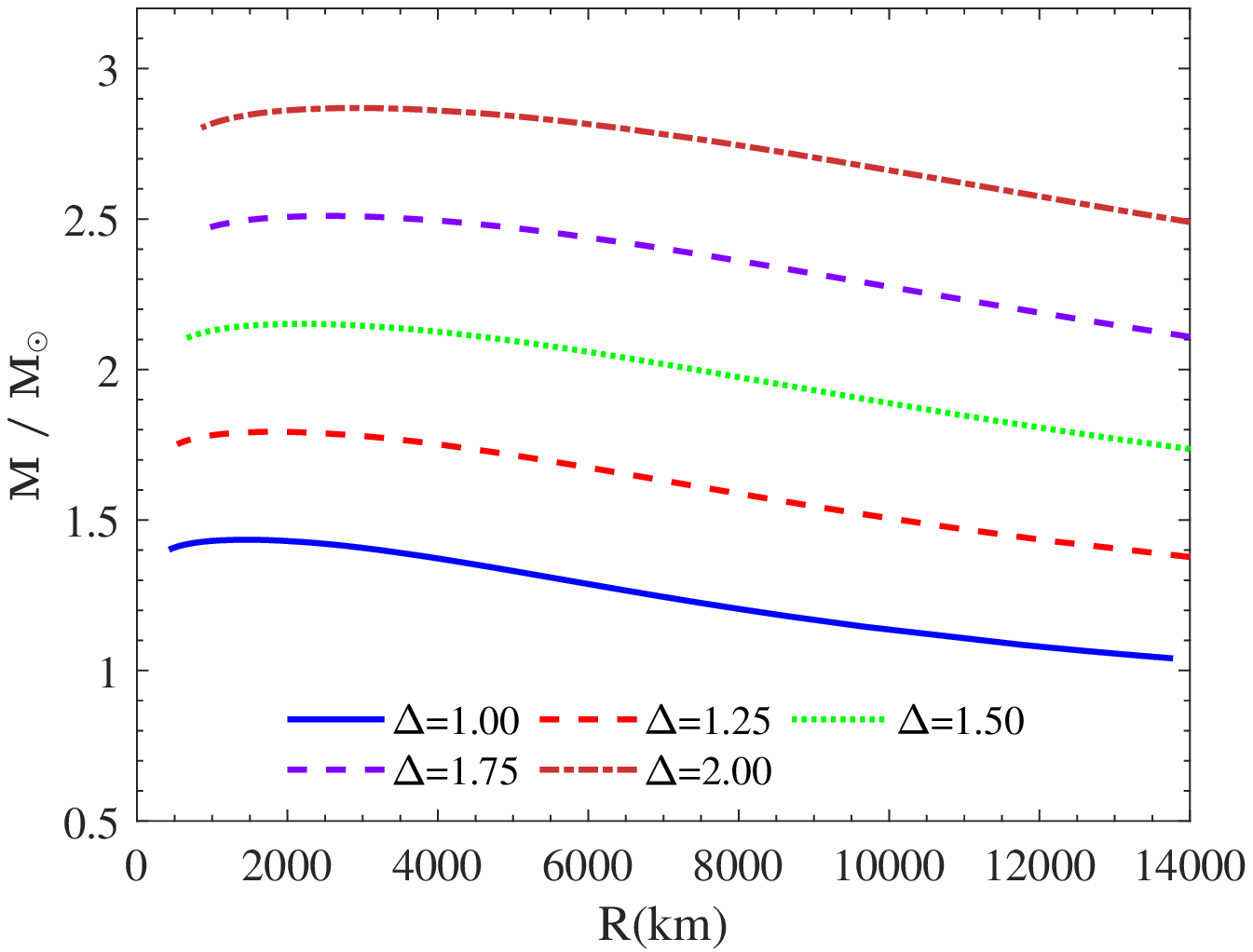}
\end{subfigure}
\begin{subfigure}[t]{0.50\textwidth}
\centering
\includegraphics[width=8.3cm,height = 6cm]{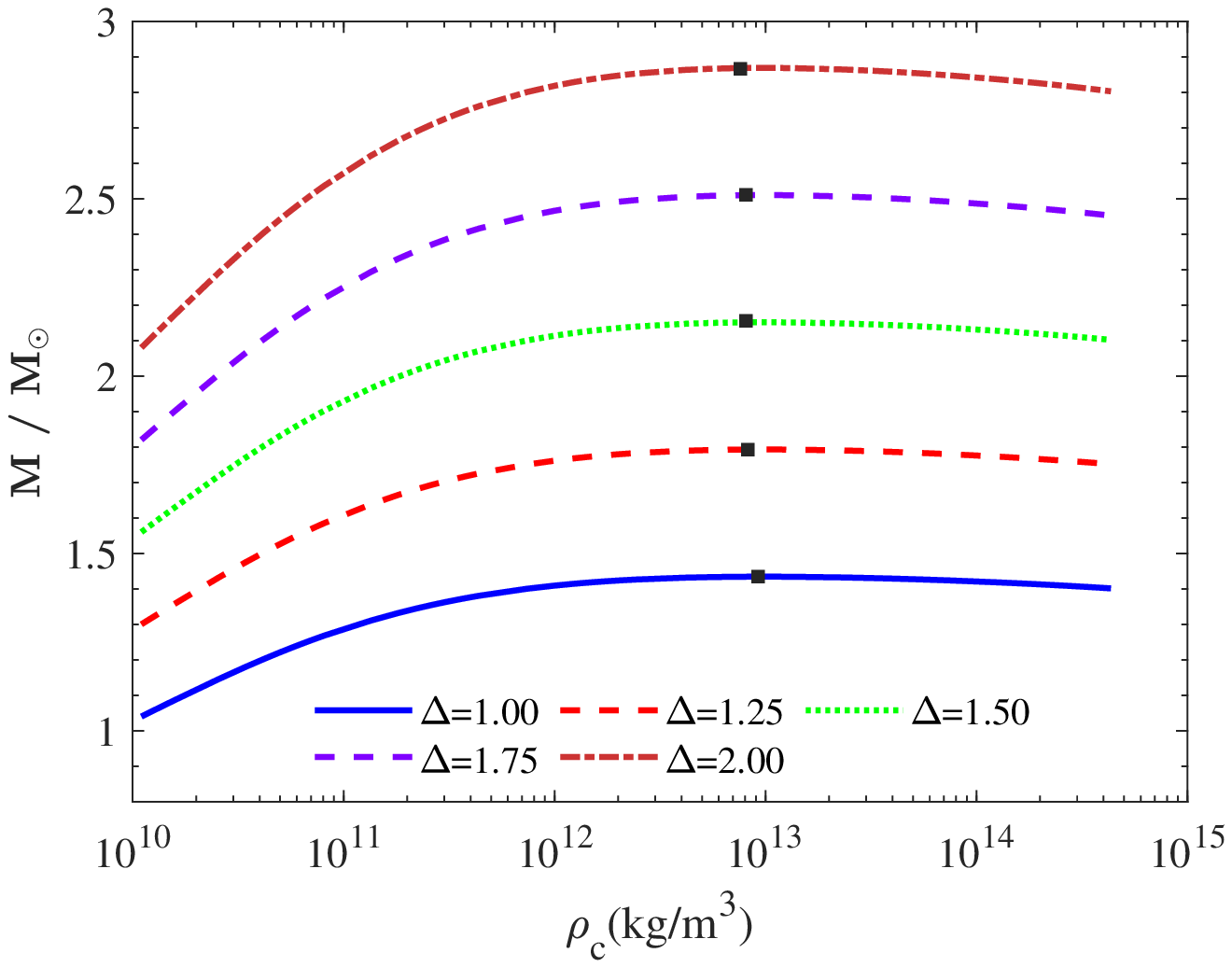}
\end{subfigure}
\caption{The mass-radius (left) and mass-central density $\rho_{c}$  relations (right) for the white stars in Rastall-Rainbow gravity for different values of $\Delta$.}
\label{Rainbow-M-R-ro}
\end{figure}
First, Fig.~\ref{M-R} shows the behavior of the mass against the radius from the center to the surface of the star for $\rho_{c} = 1\times10^{12}~$kg/m$^{3}$. Our results indicate that as the radius increases, the mass of the white dwarf increases. Furthermore, the mass obtained in the case of GR is the smallest compared to other scenarios.

 Next, we present the mass-radius relations for white dwarfs within the framework of Rastall gravity, depicted in the left panel of Fig.~\ref{Rastall-M-R-ro}. Here, we consider a fixed rainbow parameter $\Delta = 0$, while varying the parameter $\chi$ in the range of $0$ to $6\times10^{-5}$. Notably, for $\chi = 0$ (the blue solid line), the mass and radius of the white dwarf return to the results obtained in the general relativity, where the maximum mass $M = 1.41M_{\odot}$ is close to the Chandrasekhar limit. The other three curves represent the Rastall gravity case, which demonstrates that the maximum mass of white dwarfs exceeds the Chandrasekhar limit and increases with an increase in $\chi$. For example, in the case of $\chi = 4 \times 10^{-5}$, the maximum mass increases to $1.48 M_{\odot}$. The right panel of Fig.~\ref{Rastall-M-R-ro} shows the mass-center density relations, and the black dots in the figure mark the position of the maximum mass. According to the Harrison-Zeldovich Novikov criterion, $d M / d \epsilon_{c}>0$ corresponds to a stable configuration, while $d M / d \epsilon_{c}<0$ is unstable one. The results show that in the scenario where $\chi < 6\times10 ^ {-5}$, the mass of the white dwarf increases with the center density until reaching the maximum mass, indicating a state of stability. Beyond this point, the mass exhibits a decreasing trend with the center density. It is notable  that, for $\chi = 6\times10 ^ {-5}$, the mass decreases monotonically with the central density, implying that these white dwarf models are unstable. Consequently, the Chandrasekhar limit in Rastall gravity is estimated to be $M \approx 1.51 M_{\odot}$, with detailed data provided in Table~\ref{Rastall}.

\begin{figure}[t!]
\centering
\begin{subfigure}[t]{0.7\textwidth}
\centering
\includegraphics[width=11cm,height = 7cm]{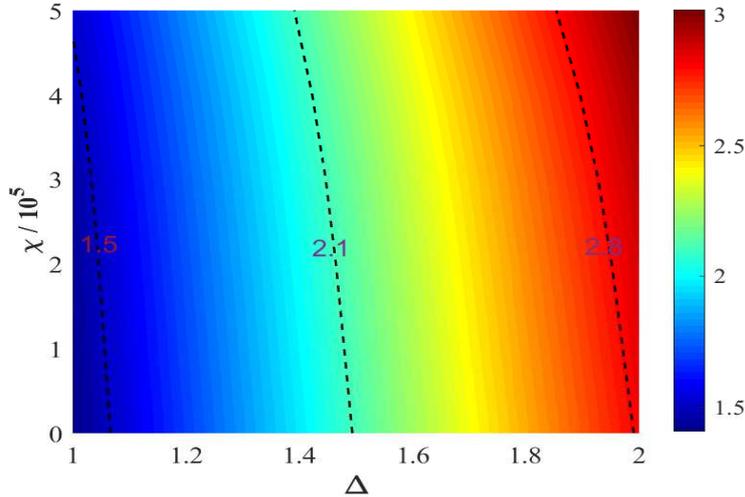}
\end{subfigure}
\caption{$\Delta-\chi$ plane for $equi-M_{max}$ contours.}
\label{Rastall-Rainbow}
\end{figure}

Finally, we investigate the effect of the rainbow parameter $\Delta$ on the properties  of white dwarfs, and the results are shown in Fig.~\ref{Rainbow-M-R-ro}. Here we fix $\chi = 2\times10^{-5}$ and vary the value of $\Delta$. The results indicate that when considering the effect of the Rainbow parameter $\Delta$, the maximum mass of white dwarfs is larger compared to that in Rastall gravity. It can be seen that super-Chandrasekhar white dwarfs can be obtained for some specific values of the Rainbow parameters.  For example, in the case of $\Delta = 1.5$, the maximum mass of the white dwarf is $2.14M_{\odot}$. Meanwhile, considering the upper limit of the super-Chandrasekhar white dwarf mass as $2.8M_{\odot}$, we can constrain the value of the rainbow parameter to $\Delta_{upper} = 2$. In this scenario, the maximum mass reaches $2.86M_{\odot}$. For more detailed analysis, Fig.~\ref{Rastall-Rainbow} shows the $equi-M_{max}$ contours in $\Delta-\chi$ plane. It can be observed that larger values of  $\chi$ and $\Delta$ facilitate the attainment of super-Chandrasekhar white dwarfs, and $\Delta_{upper}$ exhibiting a decrease as the Rastall parameter $\chi$ increases.

\begin{table}[tbp]
\centering
\begin{tabular}{|c|c|c|c|c|}
\hline
$\chi$&$M_{max}~(M_{\odot})$&R~(km)& $\sigma~(10^{-2})$ & $z_{s}~(10^{-2})$\\
\hline
  $0.0\times10^{-5}$        & 1.41 & 898   & 0.46 & 0.23\\[0.03cm]
     $1.0\times10^{-5}$        & 1.42 & 1131  &  0.37 & 0.19 \\[0.03cm]
     $2.0\times10^{-5}$        & 1.43 & 1508  & 0.27 & 0.14 \\[0.03cm]
     $3.0\times10^{-5}$        & 1.45 & 2166  & 0.19 & 0.10\\[0.03cm]
     $4.0\times10^{-5}$        & 1.48 & 3387  & 0.12 & 0.06 \\[0.03cm]
     $5.0\times10^{-5}$        & 1.51 & 6908  & 0.06 & 0.03\\[0.03cm]
\hline
\end{tabular}
\caption{\label{Rastall}Structure properties of white dwarfs for different values of $\chi$. Here $\Delta = 0$.}
\end{table}
\begin{table}[tbp]
\centering
\begin{tabular}{|c|c|c|c|c|}
\hline
$\Delta$&$M_{max}~(M_{\odot})$&R~(km)& $\sigma~(10^{-2})$ & $z_{s}~(10^{-2})$\\
\hline
         1.00         & 1.43 & 1508  & 0.27 & 0.14\\[0.03cm]
         1.25         & 1.79 & 1885  & 0.28 & 0.14 \\[0.03cm]
         1.50         & 2.14 & 2263  & 0.28 & 0.14 \\[0.03cm]
         1.75         & 2.50 & 2639 &  0.28  & 0.14\\[0.03cm]
         2.00         & 2.86 & 3016  & 0.28 &  0.14 \\[0.03cm]
\hline
\end{tabular}
\caption{\label{Rainbow}Structure properties of white dwarfs for different values of $\Delta$. Here $ \chi = 2.0\times10^{-5}$.}
\end{table}
\section{Structure properties of white dwarfs}
\subsection{ Gravitational redshift and compactness}
The gravitational redshift provides a direct measurement of the gravitational field strength near the compact star, aiding in a understanding of gravitational properties under extreme conditions. Moreover, it serves as a significant probe for investigating the EoS of compact stars, and its expression is
\begin{eqnarray}
z_{s}=\frac{1}{\sqrt{1-\sigma}}-1~.\label{redshift}
\end{eqnarray}
where $\sigma$ is the compactness of compact stars, which is defined as
\begin{eqnarray}
\sigma=\frac{2GM}{c^{2}R}~,\label{compactness}
\end{eqnarray}
The compactness and gravitational redshift of white dwarfs are given in the last two columns of Table~\ref{Rastall} and \ref{Rainbow}.  As can be seen from the above two tables, the gravitational redshift and compactness decrease with the increase of parameter $\chi$, while the rainbow parameter $\Delta$ has no significant impact on these properties.
\subsection{Dynamical stability}
\begin{figure}[h!]
\centering
\begin{subfigure}[t]{0.49\textwidth}
\centering
\includegraphics[width=8.3cm,height = 6.5cm]{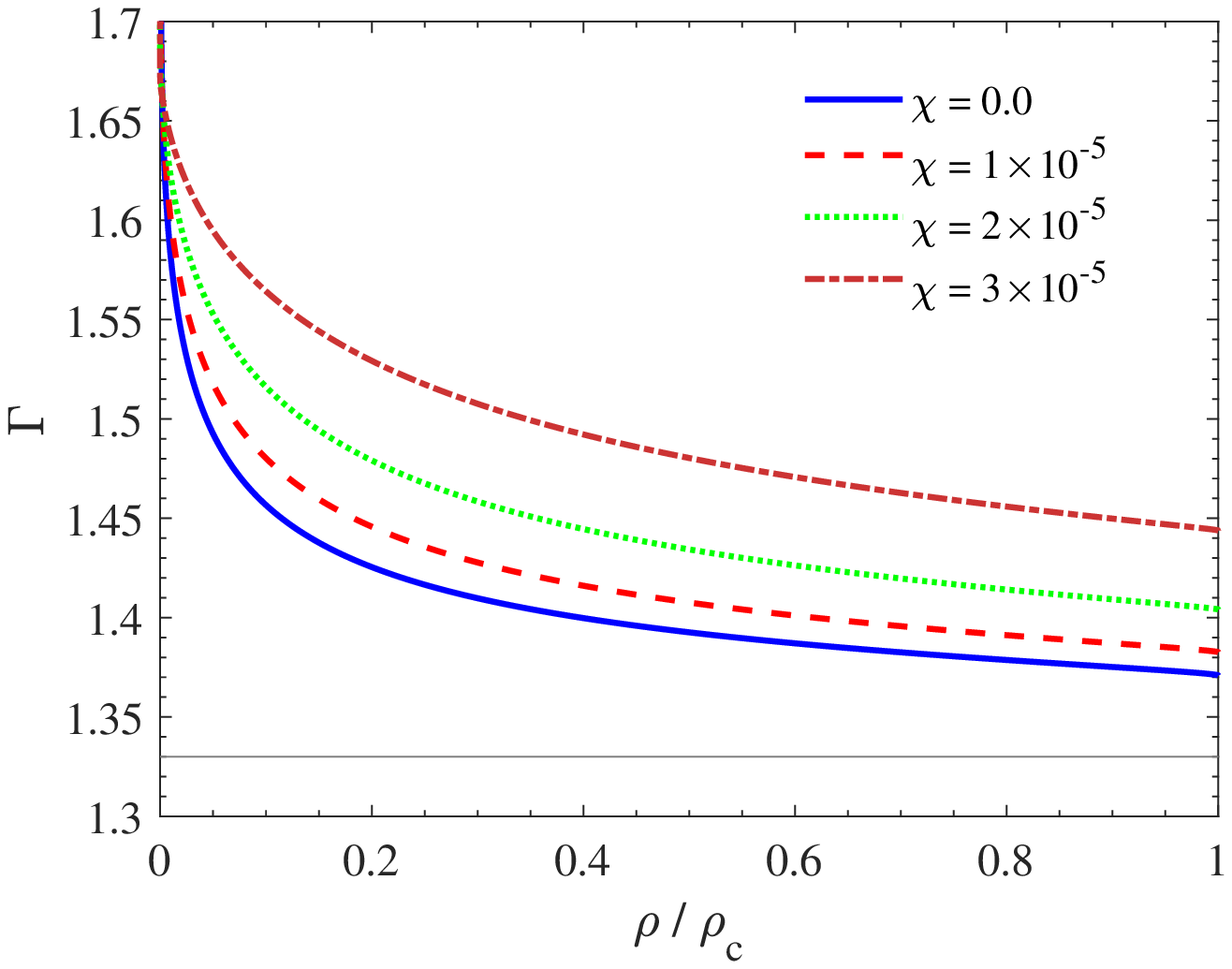}
\end{subfigure}
\begin{subfigure}[t]{0.5\textwidth}
\centering
\includegraphics[width=8.3cm,height = 6.5cm]{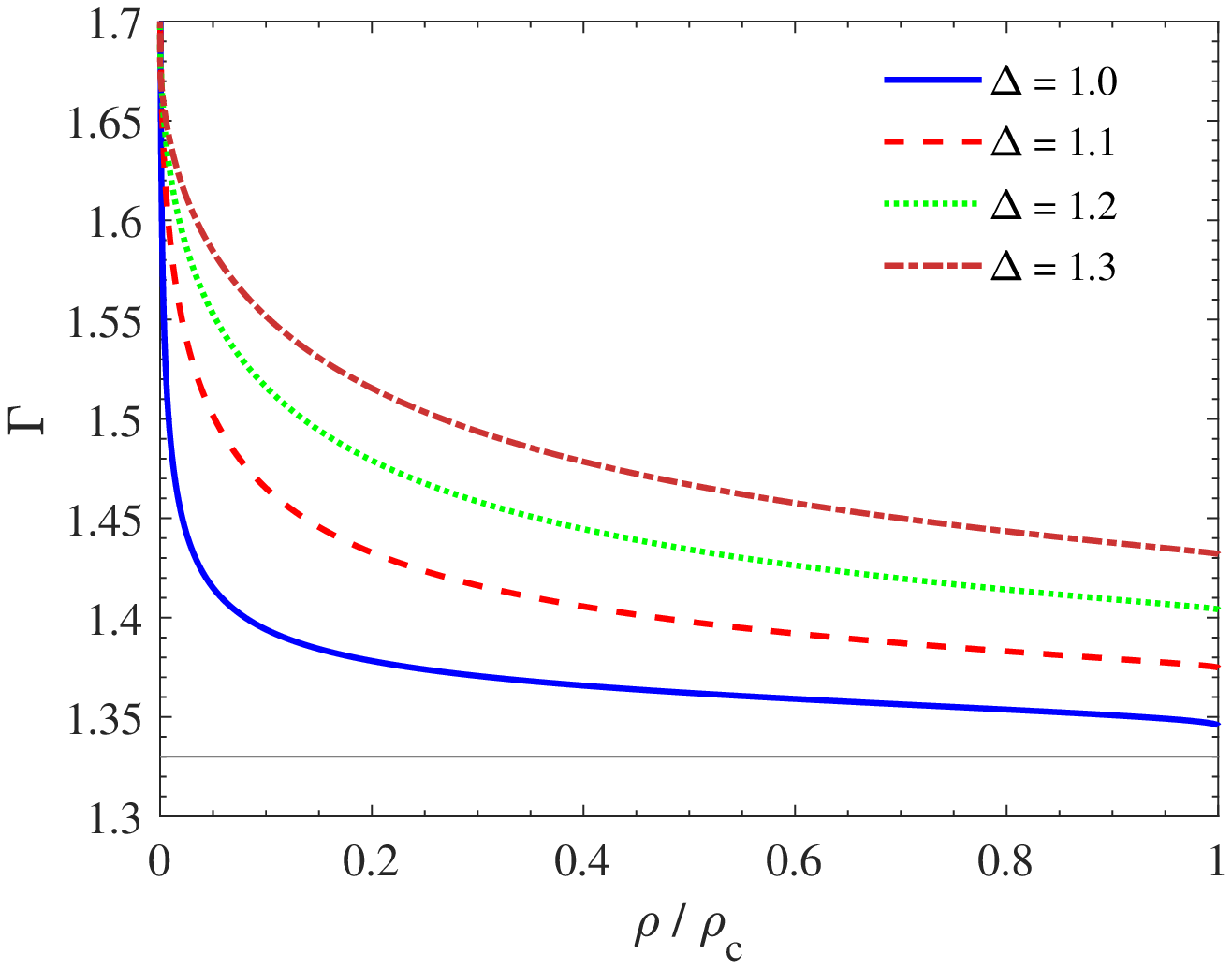}
\end{subfigure}
\caption{Adiabatic index - density relations for different values of $\chi$ and $\Delta$. Here the mass of white dwarfs is $1.4 M_{\odot}$.}
\label{adiabatic-index}
\end{figure}

The adiabatic index $\Gamma$, defined as the ratio of two specific heats, plays a crucial role in assessing the stability of white dwarfs. Chandrasekhar first studied the dynamical stability of stellar models against infinitesimal radial adiabatic perturbation~\cite{Chandrasekhar1964}. Heintzmann and Hillebrandt conducted calculations~\cite{Heintzmann1975} and established that for compact objects to be dynamically stable at any position, the adiabatic index must exceed $\frac{4}{3}$. The expression for $\Gamma$ can be written as
\begin{eqnarray}
\Gamma=\Big(\frac{c^{2}\rho +P}{c^{2}P}\Big)\frac{d P}{d\rho}~,\label{adiabatic}
\end{eqnarray}
Fig.~\ref{adiabatic-index} shows the adiabatic index versus the density for white stars in Rastall-Raindow gravity. The grey solid line represents the  critical adiabatic index, $\Gamma=1.33$. The result shows that for the different values of the parameters $\chi$ and $\Delta$ are over 1.33, which means that these white dwarf models are dynamically stable.
\section{Summary}
Super-Chandrasekhar white dwarfs are compact stellar remnants that exceed the standard mass limit. The supernova explosions triggered by these unusually massive white dwarfs exhibit distinct spectral and luminosity characteristics, making them significant for studying their properties.  In Refs~\cite{Das2015,Jing2016,Banerjee2017,Carvalho17,Kalita2018,Panah2019,Liu2019,Rocha2020,Wojnar2021,Kalita2021,Kalita2022}, authors obtained massive white dwarfs by considering various modified gravitational theories. Motivated by this, we investigate white dwarfs in Rastall-Rainbow gravity.

We numerically solve the modified TOV equation with Chandrasekhar EoS, resulting in the derivation of the mass-radius relations for white dwarfs under varying parameters of $\chi$ and $\Delta$. It was demostrated that the parameters of Rastall-Rainbow gravity  have significant effects on the structure of white dwarfs. In the realm of Rastall gravity, when $\chi>1\times10^{-5}$, the maximum mass of white dwarfs exhibits a substantial deviation from the predictions of GR, with a noticeable upward trend as $\chi$ increases. For the case of $\chi=4\times10^{-5}$, the maximum mass increases to $1.48 M_{\odot}$, which is larger than the Chandrasekhar limit.  Based on the fact that the structure of white dwarfs is unstable at $\chi>6\times10^{-5}$, we can conclude that the Chandrasekhar  limit in Rastall gravity is $M_{max}=1.51 M_{\odot}$. Futhermore, we fix $\chi = 2\times10^{-5}$ and vary the value of rainbow parameter $\Delta$. the results show that super-Chandrasekhar white dwarfs can be obtained by increasing the value of rainbow parameter $\Delta$. We can constrain $\Delta < 2$ due to that the upper limit of the super-Chandrasekhar white dwarf's mass is $2.8M_{\odot}$.

Next, we investigate the gravitational redshift and compactness of white dwarfs and find that they decrease with the increase of the parameter $\chi$, but have no significant dependence on the rainbow parameter $\Delta$. Finally, we confirm that these white dwarf models are dynamically stable by analyzing the adiabatic index.

\acknowledgments
We would like to thank Prof. Remo Riffini for helpful discussions. This work was supported by the National Natural Science Foundation of China (Grant Nos. 11973025 and 12247157),  and Scientific Research Fund of Hunan Provincial Education Department (Grant No. 22B0446).


\end{document}